\documentclass[%
reprint,
superscriptaddress,
showpacs,preprintnumbers,
 amsmath,amssymb,
 aps,
pra,
twocolumn
]{revtex4}

\usepackage{graphicx}
\usepackage{dcolumn}
\usepackage{bm}
\usepackage{color} 
\def\la{{\langle}}
\def\ra{{\rangle}}

\newcommand{\beq}{\begin{equation}}
\newcommand{\eeq}{\end{equation}}
\newcommand{\beqa}{\begin{eqnarray}}
\newcommand{\eeqa}{\end{eqnarray}}

\begin{document}
\title{Fast population transfer engineering of three-level systems}

\author{Xi Chen}

\affiliation{Departamento de Qu\'{\i}mica-F\'{\i}sica, UPV-EHU, Apdo 644, 48080 Bilbao, Spain}

\affiliation{Department of Physics, Shanghai University, 200444 Shanghai, People's Republic of China}

\author{J. G. Muga}

\affiliation{Departamento de Qu\'{\i}mica-F\'{\i}sica, UPV-EHU, Apdo 644, 48080 Bilbao, Spain}

\affiliation{Department of Physics, Shanghai University, 200444 Shanghai, People's Republic of China}

\begin{abstract}
We design, by invariant-based inverse engineering, resonant laser pulses to perform
fast population transfers in three-level systems.
The efficiency and laser intensities are examined for different
protocols.
The energy cost to improve the
fidelity is quantified. The laser intensities can be reduced by
populating the intermediate state and by multi-mode driving.
\end{abstract}
\pacs{32.80.Xx, 32.80.Qk, 33.80.Be}

\maketitle
%
%
%
%
\section{Introduction}
The laser control of internal state preparation and dynamics is of importance in atomic and molecular physics
for applications such as metrology, interferometry, quantum information processing and driving of chemical reactions  \cite{Allen,Vitanov-Rev1,Bergmann,Kral,Molmer}.
In two- or three-level systems, resonant pulses, rapid adiabatic passage (RAP),  stimulated Raman adiabatic passage (STIRAP), and their variants have been widely used to perform population transfers  \cite{Vitanov-Rev1,Bergmann,Kral}.
Generally, resonant pulses may be fast, but are highly sensitive to the deviations of pulse areas and exact resonances, whereas
the adiabatic passage techniques are robust versus  variations in experimental parameters but slow.
To combine the advantages of resonant pulses and adiabatic techniques
and achieve fast and high-fidelity quantum state control,
some alternative approaches, like composite pulses \cite{Levitt,VitanovRRA11,VitanovPRL11} and optimal control theory \cite{Boscain,Sugny,Vitanov09}, have been proposed.
Several recent works on ``shortcuts to adiabaticity'' have been also devoted to internal state population transfer and control \cite{Rice,Berry09,Chen10b,Oliver,ChenPRA,nonHermitian,WanYidun,Sara11}.
The shortcut techniques include counter-diabatic control protocols \cite{Rice} or, equivalently, transitionless quantum driving \cite{Berry09,Chen10b,Oliver},
``fast-forward" scaling \cite{Masuda}, and inverse engineering based on Lewis-Riesenfeld invariants \cite{bec,Chen}.
These methods are in fact strongly related, and even potentially equivalent
\cite{ChenPRA,ErikFF}. However, they provide in general different shortcuts
\cite{ChenPRA,Sara11}.

In this paper, we apply invariant-based engineering to realize  fast and robust population transfers in three-level systems.
This method has been applied in trap expansions \cite{bec,Chen,energy,Li,Nice,Nice2}, rotations \cite{Nice3},
atom transport \cite{transport,opttransport,transport2}, mechanical oscillators
\cite{Wu}, or many-body systems \cite{Adol,Onofrio}.
In a three-level system as the one depicted in Fig. \ref{fig1}, STIRAP allows to transfer the population adiabatically from the initial state $|1\ra$ to the target state $|3 \ra$. To speed up the process, a fast-driving counterdiabatic field connecting levels $|1\ra$ and $|3\ra$ may be used  \cite{ShoreOC,Chen10b}.
In general, though, this implies a weak magnetic dipole transition, which limits
the ability of the counterdiabatic field to shorten the times \cite{ShoreOC,Chen10b}.
This will be solved by invariant-based engineering, which  provides
alternative shortcuts without
coupling directly levels $|1\ra$ and $|3\ra$.

%
%
%
%
\section{Invariant dynamics}
\label{invariant}
%
%
%
%
%
\begin{figure}[]
\scalebox{0.25}[0.25]{
\includegraphics{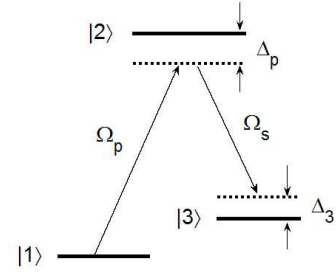}}
\caption{Level scheme of STIRAP for a $\Lambda$ level configuration.
$\Omega_p$ and $\Omega_s$ are the Rabi frequencies for the interactions with the pump and
Stokes fields respectively, and $\Delta_p$ and $\Delta_3$ are the detunings from the resonances.}
\label{fig1}
\end{figure}
%
To perform  STIRAP, the Hamiltonian within the rotating wave approximation (RWA) reads
\beqa
\label{gernal}
H_0 (t)= \frac{\hbar}{2} \left(\begin{array}{ccc} 0 & \Omega_{p}(t) & 0 \\ \Omega_{p}(t) & 2 \Delta_{p} & \Omega_{s}(t) \\ 0 & \Omega_{s}(t) & 2\Delta_{3}
\end{array}\right),
\eeqa
where, as shown in Fig. \ref{fig1}, the coupling strengths between the states are determined by $\Omega_{s}(t)$ and $\Omega_{p}(t)$, which describe the interactions with the pump and Stokes fields, and the detunings from resonance are defined as $\Delta_p = (E_2-E_1)/\hbar - \omega_p$,
$\Delta_s=(E_2-E_3)/\hbar -\omega_s$, and $\Delta_3=\Delta_s-\Delta_p$,
where $\omega_s$ and $\omega_p$ are
the laser (angular) frequencies, and the $E_j$, $j=1,2,3$, the bare-basis-state energies.

We consider the so called ``one-photon resonance'' case, $\Delta_p=\Delta_3=0$
to simplify the Hamiltonian as
\beqa
\label{H0}
H (t)= \frac{\hbar}{2} \left(\begin{array}{ccc} 0 & \Omega_{p}(t) & 0 \\ \Omega_{p}(t) & 0 & \Omega_{s}(t) \\ 0 & \Omega_{s}(t) & 0
\end{array}\right).
\eeqa
The corresponding instantaneous eigenstates $|n\ra$, with eigenvalues
$E_{0} = 0$ and $E_{\pm} = \pm \hbar \Omega/2 $ with $\Omega =\sqrt{\Omega^2_{p} + \Omega^2_{s}}$, are
\beqa
\label{instantaneuous states}
|n_0 (t)\rangle = \left(\begin{array}{ccc} \cos \theta  \\ 0 \\ -\sin \theta \end{array}\right),~~
|n_{\pm} (t)\rangle = \frac{1}{\sqrt{2}} \left(\begin{array}{ccc} \sin \theta \\ \pm 1  \\ \cos \theta
 \end{array}\right), ~~
\eeqa
where $\tan \theta =\Omega_{p}/ \Omega_{s}$.
When the adiabatic condition,
$
|\dot{\theta}| \ll |\Omega|,
$
is satisfied, perfect population transfer from the ground state $|1 \rangle$ to
the excited state $|3 \rangle$ can be achieved adiabatically along the
dark state $|n_0 \rangle$, using the counterintuitive pulse order (Stokes before pump).
In what follows we shall use a dynamical invariant to speed up the population transfer.

To construct the dynamical invariant, the Hamiltonian in Eq. (\ref{H0}) can be rewritten as \cite{Carroll}
\beq
H(t)= \frac{\hbar}{2} (\Omega_p (t) \hat{K}_1 +\Omega_s (t) \hat{K}_2),
\eeq
where $\hat{K}_1$, $\hat{K}_2$, and $\hat{K}_3$ are angular-momentum operators for spin $1$ \cite{Carroll},
$$
\hat{K}_1= \left(\begin{array}{ccc} 0 & 1 & 0 \\ 1 & 0 & 0 \\ 0 & 0 & 0
\end{array}\right),  \hat{K}_2 = \left(\begin{array}{ccc} 0 & 0 & 0 \\ 0 & 0 & 1 \\ 0 & 1 & 0
\end{array}\right), \hat{K}_3= \left(\begin{array}{ccc} 0 & 0 & -i \\ 0 & 0 & 0 \\ i & 0  & 0
\end{array}\right),
$$
that satisfy the commutation relations
\beq
[\hat{K}_1, \hat{K}_2] = i\hat{K}_3,~ [\hat{K}_2, \hat{K}_3] = i \hat{K}_1,~ [\hat{K}_3, \hat{K}_1] = i\hat{K}_2.
\eeq
The Hamiltonian (\ref{H0}) possesses SU(2) dynamical symmetry, and an invariant $I(t)$, satisfying the condition
$d I/d t\equiv \partial I(t)/\partial t + (1/i \hbar) [I(t), H(t)] =0$, can be  constructed as \cite{Liang,ChenPRA}
\beqa
\label{I}
I(t) &=& \frac{\hbar}{2} \Omega_0 (\cos\gamma \sin\beta \hat{K}_1 +\cos\gamma \cos\beta \hat{K}_2 + \sin \gamma \hat{K}_3),
\nonumber \\
&=& \frac{\hbar}{2}  \Omega_0 \left(\begin{array}{ccc} 0 & \cos\gamma \sin\beta & - i\sin \gamma  \\ \cos\gamma \sin\beta & 0 & \cos\gamma \cos\beta \\ i \sin \gamma  & \cos\gamma \cos\beta & 0
\end{array}\right),
\eeqa
where $\Omega_0$ is an arbitrary constant with units of frequency to keep $I(t)$ with dimensions of energy, and the time-dependent auxiliary parameters $\gamma$ and
$\beta$ satisfy the equations
\beqa
\label{gamma}
\dot{\gamma} &=& \frac{1}{2} (\Omega_p \cos{\beta}-\Omega_s \sin{\beta} ),
\eeqa
\beqa
\label{beta}
\dot{\beta} &=& \frac{1}{2}  \tan{\gamma} (\Omega_s \cos{\beta}+\Omega_p \sin{\beta}).
\eeqa
The eigenstates of the invariant $I(t)$, satisfying $I(t) |\phi_n (t)\rangle = \lambda_n |\phi_n (t)\rangle$, (we use the labels $n=0,\pm$) are
\beqa
\label{eigenstate inv-1}
|\phi_{0} (t) \rangle = \left(\begin{array}{ccc} \cos \gamma \cos \beta  \\ - i \sin{\gamma} \\ -\cos{\gamma} \sin{\beta}  \end{array}\right),
\eeqa
and
\beqa
\label{eigenstate inv-2}
|\phi_{\pm} (t) \rangle = \frac{1}{\sqrt{2}} \left(\begin{array}{ccc} \sin \gamma \cos \beta \pm i \sin\beta \\ i \cos\gamma \\ -\sin \gamma \sin \beta \pm
i \cos \beta
\end{array}\right),
\eeqa
which correspond to the eigenvalues $\lambda_0 =0$ and $\lambda_{\pm} = \pm 1$. According to Lewis-Riesenfeld theory \cite{LR},
the solution of the Schr\"{o}dinger equation, $i \hbar \partial_t \Psi = H \Psi $, is a superposition
of orthonormal ``dynamical modes", $\Psi (t) = \sum_n C_n e^{i \alpha_n} |\phi_{n} (t) \ra$ \cite{LR}, where each $C_n$ is a time-independent amplitude
and $\alpha_n$ is the Lewis-Riesenfeld phase,
\beq
\label{LR phase}
\alpha_n (t) =\frac{1}{\hbar} \int^t_0 \langle \phi_n(t') | i\hbar \frac{\partial }{\partial t'} - H(t')| \phi_n(t') \rangle dt'.
\eeq
In our case  $\alpha_0 = 0$, whereas
$$
\alpha_{\pm} = \mp \int^t_0 \left[\dot{\beta} \sin \gamma + \frac{1}{2}\left(\Omega_p \sin\beta +\Omega_s \cos\beta\right)\cos\gamma\right] dt'.
$$
The dot represents here a time derivative.
%
%
%
%
\section{Inverse engineering and fast population transfer}
\label{inverse engineering}
Beginning with the Hamiltonian in Eq. (\ref{H0}), we shall apply invariant-based inverse engineering to design $\Omega_s$ and $\Omega_p$.
We firstly assume that $\Omega_s$ and $\Omega_p$ are unknown functions to be determined, from (\ref{gamma}) and (\ref{beta}),
\beqa
\label{omega-s}
\Omega_s &=& 2( \dot{\beta} \cot \gamma \cos \beta-\dot{\gamma} \sin \beta), ~~~~
\\
\label{omega-p}
\Omega_p &=& 2( \dot{\beta} \cot\gamma \sin\beta+\dot{\gamma} \cos \beta).
\eeqa
Once the appropriate boundary conditions for
$\gamma$ and $\beta$ are fixed, we are ready to choose some ansatz, for example, a polynomial or some other function with enough
free parameters, to construct $\Omega_s$ and $\Omega_p$.

Our Hamiltonian $H(t)$, Eq. (\ref{H0}), should drive the initial state $|1 \rangle$
to the target state $|3 \rangle$, up to a phase factor, along the invariant eigenstate $|\phi_{0} (t) \ra$ in a given time $t_f$.
We therefore write down the boundary conditions for $\gamma$ and $\beta$, based on Eq. (\ref{eigenstate inv-1}),
\beqa
\label{BC-gamma}
\gamma (0)= 0, ~~ \gamma (t_f)=0,
\\
\label{BC-beta}
\beta (0) = 0,  ~~\beta (t_f) = \pi/2.
\eeqa
In general, $H (t)$ does not commute with the invariant $I(t)$, 
%
%
which means they do not have common eigenstates.
To achieve fast adiabatic-like passage (i.e., not really adiabatic all along but leading to the same final result), one may impose boundary conditions to satisfy $[H (0), I(0)]=0$
and $[H(t_f), I(t_f)]=0$, which give $\Omega_p(0)=0$ and $\Omega_s(t_f)=0$. Using  Eqs. (\ref{omega-s})-(\ref{BC-beta}) this implies the additional boundary conditions
\beqa
\label{BC-dotgamma}
\dot{\gamma} (0)=0, ~~ \dot{\gamma} (t_f) =0.
\eeqa
The set of conditions in Eqs. (\ref{BC-gamma})-(\ref{BC-dotgamma}) guarantee fast adiabatic-like population transfer.
Now we are ready to apply inverse engineering and design different protocols.

\begin{figure}[]
\scalebox{0.55}[0.55]{\includegraphics{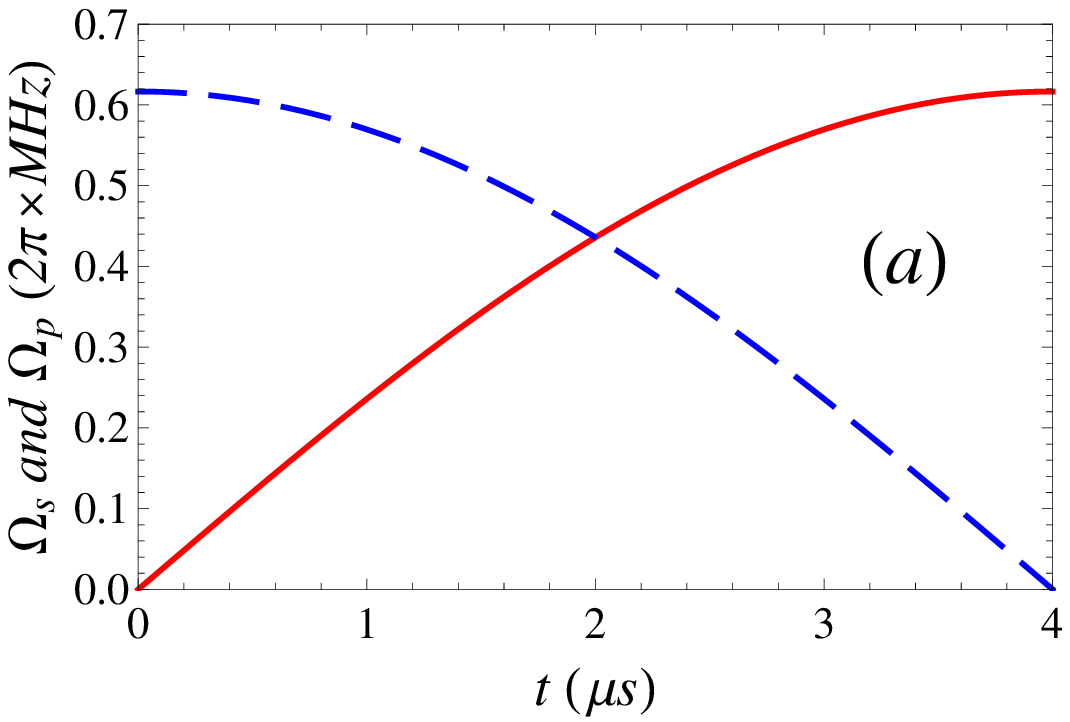}}
\scalebox{0.55}[0.55]{\includegraphics{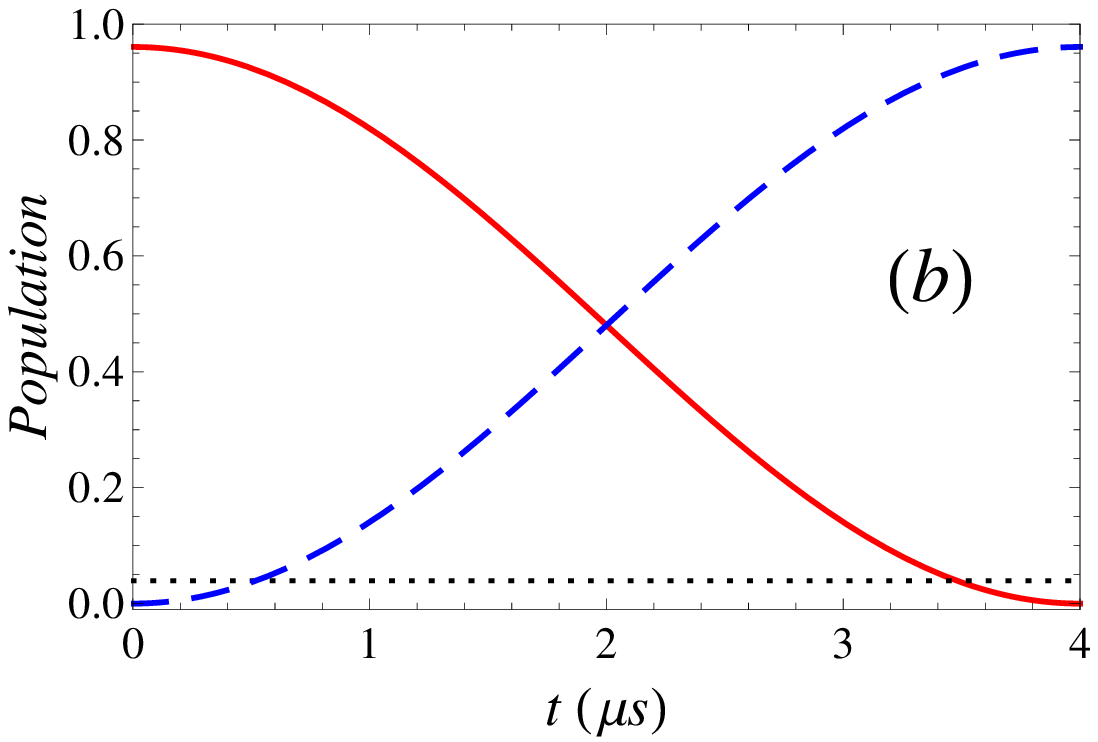}}
\caption{(Color online) (a) Time evolution of Rabi frequencies, $\Omega_p$ (solid red) and $\Omega_s$ (dashed blue) for $\gamma (t)= \epsilon$ and $\beta (t) = \pi t/ 2 t_f$. (b) Time evolution of the corresponding populations of levels $1$ (solid red), $2$ (dashed blue), and $3$ (dotted black).
Parameters: $t_f = 4 \mu s$, $\epsilon=0.2$.}
\label{example1}
\end{figure}

{\textit{Protocol 1}}- In the first example, we set the boundary conditions for $\gamma$ and $\beta$ as follows:
\beqa
\label{boundarycondition-1a}
\gamma (0)= \epsilon, ~~ \dot{\gamma} (0)=0, ~~ \gamma (t_f)=\epsilon, ~~ \dot{\gamma} (t_f) =0,
\\
\label{boundarycondition-1b}
\beta (0) = 0,  ~~\beta (t_f) = \pi/2.~~~~~~~~~~~~
\eeqa
Note that we have introduced a small value $\epsilon$ for $\gamma$,
as an exact zero value implies infinite Rabi-frequencies according to Eqs. (\ref{omega-s}) and (\ref{omega-p}).
With these boundary conditions, we can simply choose
\beqa
\label{function-1}
\gamma (t)= \epsilon,~~  \beta (t) = \pi t/ 2 t_f,
\eeqa
which provide
\beqa
\label{Omegas-1}
\Omega_s (t) &=& (\pi/t_f) \cot{\epsilon} \cos{(\pi t/ 2 t_f)},
\eeqa
\beqa
\label{Omegap-1}
\Omega_p (t) &=& (\pi/t_f) \cot{\epsilon} \sin{(\pi t/ 2 t_f)}.
\eeqa
Fig. \ref{example1} shows the time evolution of Rabi frequencies and corresponding population transfer for $\Psi(t)$ with
initial and final states $|\phi_0 (0) \ra$ and $|\phi_0(t_f) \ra$. 
We take $|-3\ra = (0,0,-1)^{T}$ as the target state, which corresponds to $|\phi_0(t_f)\ra$ for
the ideal conditions $\gamma(t_f)=0$ and $\beta(t_f)=\pi/2$.
(Note that for $\epsilon\ne 0$ the initial state is not exactly $|1\ra$.
In Protocol 3, below, we shall examine the case $|\Psi(0)\ra=|1\ra$.)
The final fidelity with the target state is
\beq
\label{fidelity-1}
F \equiv \la -3 | \Psi (t_f) \ra = \cos\epsilon.
\eeq
From Eqs. (\ref{Omegas-1})-(\ref{fidelity-1}), we find
\beq
\frac{\partial \Omega_s}{\partial \epsilon} = \frac{\partial \Omega_p}{\partial \epsilon} = - \frac{\pi \cos{(\pi t/ 2 t_f)}}{t_f \sin^2{\epsilon}} \sim - \frac{1}{\epsilon^2},
\eeq
and
\beq
\frac{\partial F}{\partial \epsilon} = - \sin{\epsilon} \sim - \epsilon,
\eeq
respectively. In other words, the fidelity varies smoothly with $\epsilon$, whereas the Rabi frequencies decrease dramatically when increasing
$\epsilon$. This provides the possibility to achieve a desired fidelity with relatively small Rabi frequencies.

Improving the fidelity or shortening $t_f$ implies increasing the Rabi frequencies.
Note the behavior of the time-averaged frequency,
\beq
\label{energy-0}
\overline{\Omega} \equiv \frac{1}{t_f} \int_0^{t_f}\!\sqrt{\Omega^2_s+\Omega^2_p}\, dt = \frac{\pi \cot{\epsilon}}{t_f},
\eeq
and the ``energy cost'' (time-averaged energy),
\beq
\label{energy-1}
\overline{E}/\hbar \equiv \int_0^{t_f}\!(\Omega^2_s+\Omega^2_p)\,dt = \frac{\pi^2 \cot^2{\epsilon}}{t_f}.
\eeq
%
%
%
\begin{figure}[]
\scalebox{0.55}[0.55]{\includegraphics{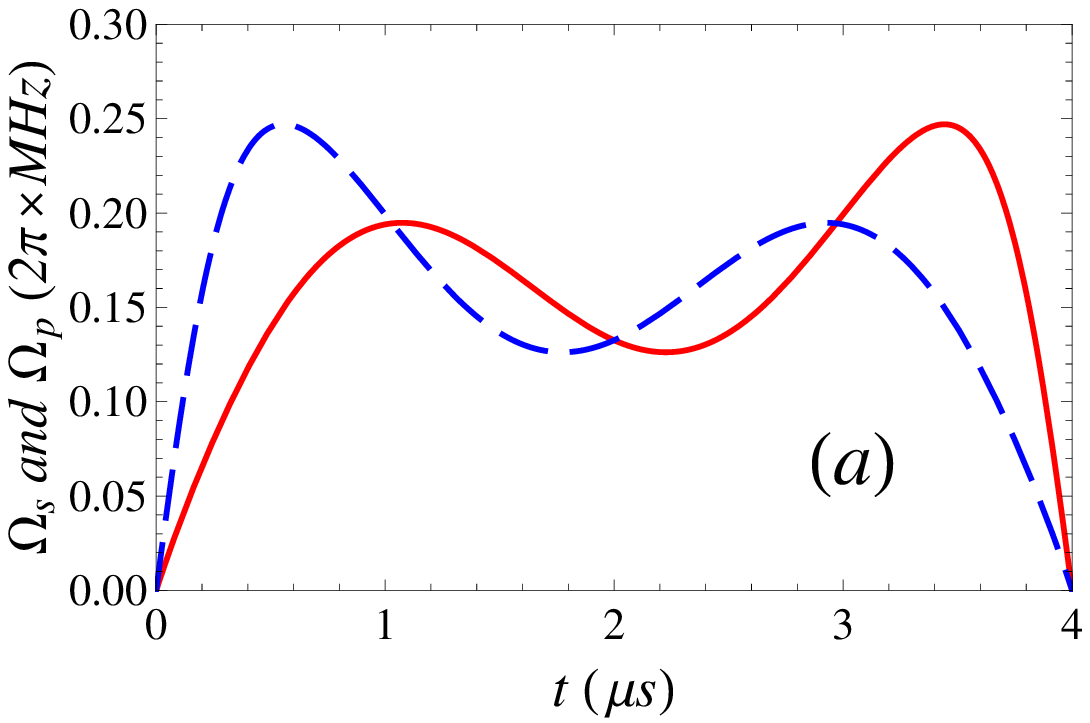}}
\scalebox{0.55}[0.55]{\includegraphics{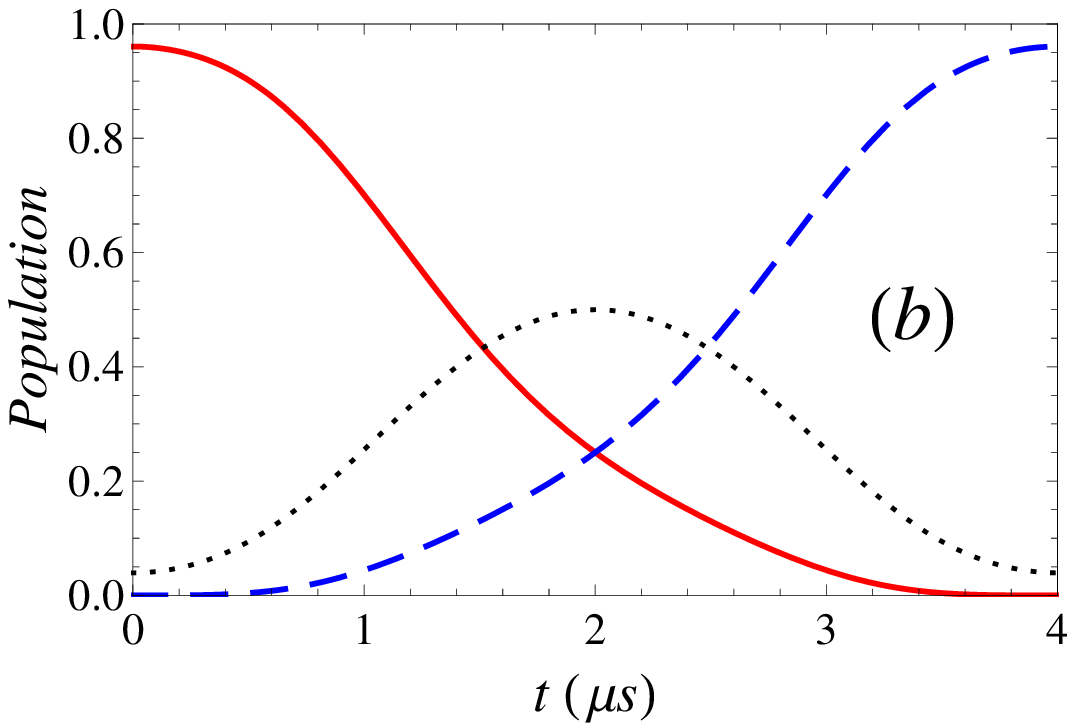}}
\caption{(Color online) (a) Time evolution of Rabi frequencies, $\Omega_p$ (solid red) and $\Omega_s$ (dashed blue), for $\gamma (t) = \sum^4_{j=0} a_j t^j$, and
$ \beta (t) = \sum^3_{j=0} b_j t^j$, with the boundary conditions (\ref{boundarycondition-2a})-(\ref{boundarycondition-2c}).
(b) Time evolution of the corresponding populations of levels $1$ (solid red), $2$ (dashed blue), and $3$ (dotted black).
Parameters: $\delta=\pi/4$, $t_f = 4 \mu s$, $\epsilon=0.2$. }
\label{example2}
\end{figure}

\textit{Protocol 2}- We design now a different protocol,
in which the intermediate state $|2\ra$ may be populated, and both pump and Stokes pulses vanish at $t=0$ and $t=t_f$. Thus, we set the following boundary conditions:
\beqa
\label{boundarycondition-2a}
\gamma (0)= \epsilon, ~~ \dot{\gamma} (0)=0, ~~ \gamma (t_f)=\epsilon, ~~ \dot{\gamma} (t_f) =0,
\\
\label{boundarycondition-2b}
\beta (0) = 0,~~\beta (t_f) = \pi/2,~~~~~~~~~~~~~
\\
\label{boundarycondition-2c}
\gamma(t_f/2)=\delta,~~\dot{\beta}(0)=0, ~~ \dot{\beta}(t_f)=0.~~~~~~~~~
\eeqa
The boundary conditions in Eqs. (\ref{boundarycondition-2a}) and (\ref{boundarycondition-2b}) are the same as before, but we add now
Eq. (\ref{boundarycondition-2c}): since the population
of the intermediate state $|2\ra$ is given by $P_2=\sin^2{\gamma}$, the condition $\gamma(t_f/2)=\delta$  sets its maximal value at $t=t_f/2$, whereas
$\dot{\beta}(0)=0$ and  $\dot{\beta}(t_f)=0$ guarantee that $\Omega_s (0) =0$ and $\Omega_p (t_f) =0$.

By assuming a polynomial ansatz, $\gamma (t) = \sum^4_{j=0} a_j t^j$ and
$\beta (t) = \sum^3_{j=0} b_j t^j$,  to interpolate at intermediate times,
we can solve the coefficients in terms of the boundary conditions. Once $\gamma(t)$ and $\beta(t)$ are fixed, we may calculate the time evolution of pulses and populations,  see e.g. Fig. \ref{example2}, where
$\delta=\pi/4$ is chosen as an example. Fig. \ref{example2} shows that the intermediate level $|2\ra$ is populated,
and the population is $1/2$ at $t=t_f/2$, because $\gamma(t_f/2) = \pi/4$.
The two examples are compared in Figs. \ref{example1} and \ref{example2}:
the laser pulse intensity is smaller when
the intermediate state $|2\ra$ is allowed to be populated.
Note that while sharing the SU(2) dynamical symmetry with two-level systems
\cite{ChenPRA}, the three-level system cannot be reduced
to a two-level system.


\begin{figure}[]
\scalebox{0.55}[0.55]{\includegraphics{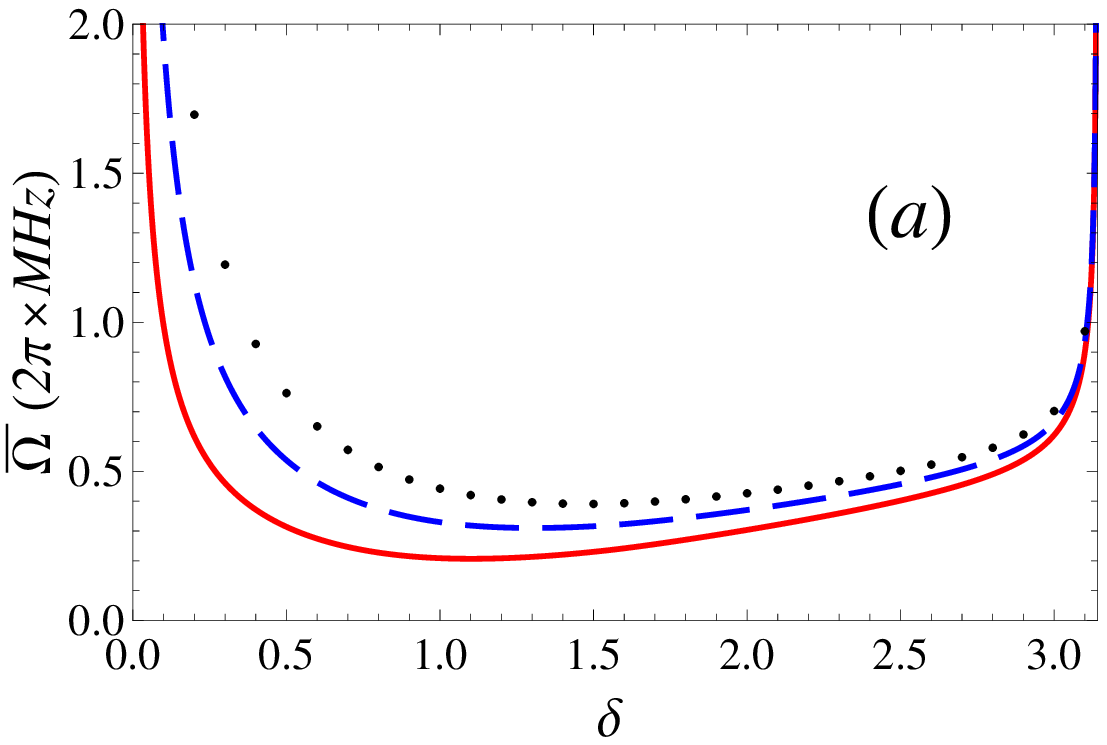}}
\scalebox{0.55}[0.55]{\includegraphics{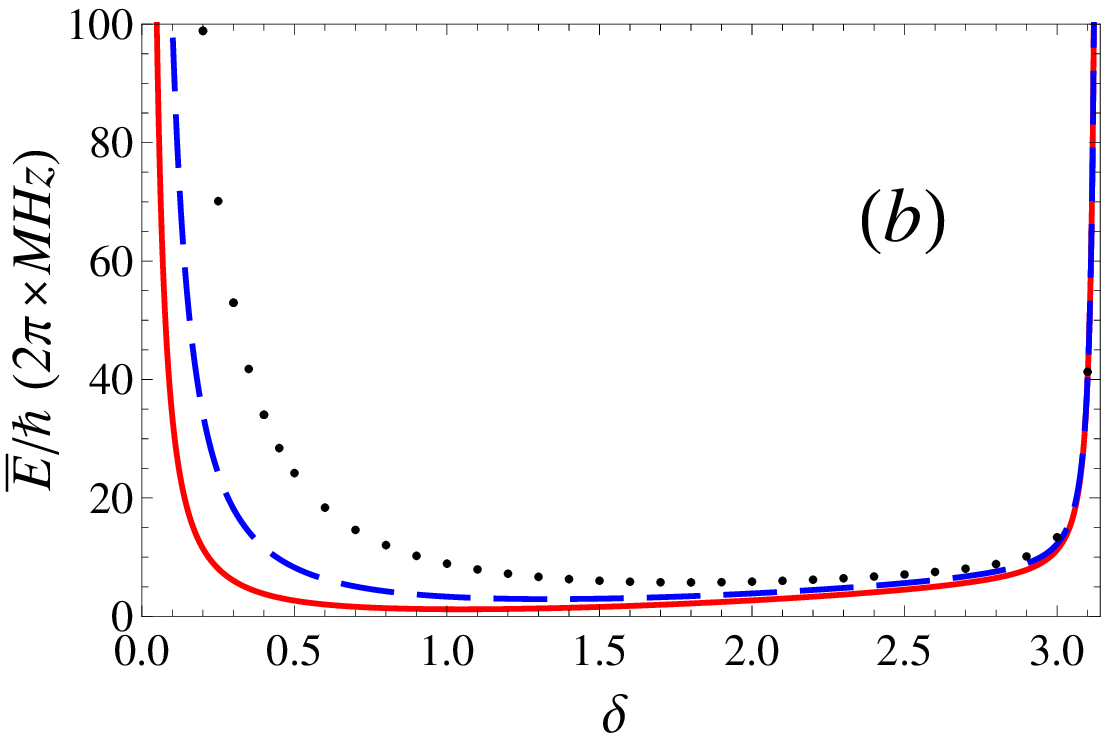}}
\caption{(Color online) Time-averaged frequency (a) and energy (b) in the second protocol as a function of $\delta$ for different values of $\epsilon$, where $\epsilon = 0.2$ (solid red),
$\epsilon= 0.02$ (dashed blue), and $\epsilon = 0.002$ (dotted black).
Other parameters are the same as in Fig. \ref{example2}.}
\label{energy}
\end{figure}

We also calculate the time-averaged Rabi frequency and energy in Fig. \ref{energy}.
Fig. \ref{energy} demonstrates that the time-averaged frequency and energy increase for a smaller $\epsilon$ as before.
When $\epsilon=0.002$ in Fig. \ref{energy}, the fidelity $F=\cos{\epsilon}$ is equal to $.999998$, which satisfies the criterion for a fault-tolerant quantum computer \cite{Molmer}.
They also decrease significantly by populating level $|2\ra$, though the
behaviors of frequency and energy are not the same.
Remarkably, Fig. \ref{energy} (a) shows that
the time-averaged frequency for each $\epsilon$ can be minimized. For the smallest  $\epsilon$ this happens when $\delta$ approaches $\pi/2$.
In this case, the intermediate state is fully populated.
The time-averaged energy is even flatter for central values of $\delta$, see Fig. \ref{energy} (b).
When the intermediate state is not populated at all,
that is, $\delta=0$ or $\delta=\pi$, both time-averaged energy and frequency increase dramatically.

In general we may combine the invariant-based method with
optimal control to optimize the protocols according to different physical criteria \cite{Li,opttransport,transport2}, for example, (time-averaged) frequency minimization or energy minimization. 
The time-optimal problem with bounded energy, and the minimum energy cost problem for fixed time
have been solved for the three-level system \cite{Boscain,Sugny}.

\textit{Protocol 3}-
Our last protocol may be considered as a variant of the first one, with the same pulses
but a different initial state. An important difference with respect to the
previous protocols is that it is based on multi-mode driving rather than on a single-mode driving. This means that the time dependent wave function
$|\Psi(t)\ra$ will
include contributions from the three eigenvectors of the invariant.

\begin{figure}[]
\scalebox{0.55}[0.55]{\includegraphics{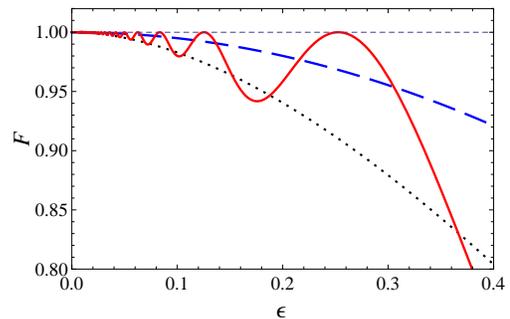}}
\caption{(Color online) Fidelity versus $\epsilon$ for the initial state $|1\ra$:  solid red and dotted black lines correspond to the examples
$1$ and $2$ in Figs. \ref{example1} and \ref{example2}. The fidelity $F= \cos{\epsilon}$ (dashed blue) for the initial
state $|\phi_0 (0)\ra$ is also shown.}
\label{fidelity}
\end{figure}

So far we have assumed that the initial state depends on $\epsilon$ through the
dependence of $|\phi_0(0)\ra$ on $\epsilon$.
Let us instead use the bare state $|1\ra$ as initial state
but keep the designed interactions $\epsilon$-dependent as before.
%
Fig. \ref{fidelity} shows the fidelity $\la -3 | \Psi (t_f) \ra$ as a function of $\epsilon$ when the initial state is $|1\ra$. The fidelity for the $\epsilon$-dependent initial state
$|\phi_0 (0)\ra$ is also shown for comparison.

\begin{figure}[]
\scalebox{0.55}[0.55]{\includegraphics{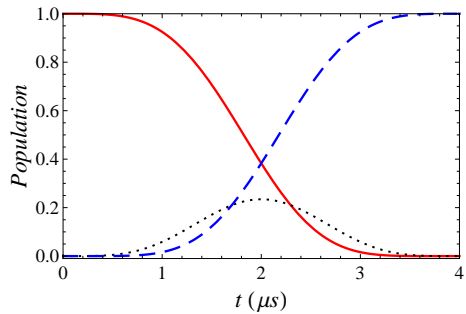}}
\caption{(Color online) Time evolution of the populations of levels $1$ (solid red), $2$ (dashed blue), and $3$ (dotted black), where
the pump and Stokes pulses are described by Eqs. (\ref{Omegas-1}) and (\ref{Omegap-1}), and the initial state is $|1\ra$.
Parameters: $t_f = 4 \mu s$, $\epsilon=0.2527$. }
\label{example3}
\end{figure}

Interestingly, the fidelity oscillates with respect to $\epsilon$ for the pump and Stokes pulses described by Eqs. (\ref{Omegas-1}) and (\ref{Omegap-1}). To analyze this in more detail, we first calculate the final state $\Psi(t_f) = \sum_n C_n e^{i \alpha_n} |\phi_n (t_f) \ra$, where $C_n = \la \phi_n (0) | 1\ra$. With the eigenvectors $|\phi_n (t)\ra$ at $t=0$ and $t_f$ we have
\beq
F \equiv \la -3 | \Psi (t_f) \ra = e^{i \alpha_{0}} \cos^2{\epsilon} + \frac{1}{2} (e^{i \alpha_{+}} +e^{i \alpha_{-}})\sin^2{\epsilon}.
\eeq
In the first protocol, the Lewis-Riesenfeld phases $\alpha_n$ are
\beq
\alpha_0=0, ~~ \alpha_{\pm} = \mp \frac{\pi}{2 \sin{\epsilon}},
\eeq
which finally gives
\beq
F = 1 - \sin^2{\epsilon} \left\{1 - \cos{\left(\frac{\pi}{2\sin{\epsilon}}\right)}\right\}.
\eeq
When the condition
\beq
\label{condition}
(\sin{\epsilon})^{-1}= 4 N,~~ (N=1,2,3...)
\eeq
is satisfied, the fidelity becomes $1$. By solving Eq. (\ref{condition}), we
get $\epsilon = 0.2527$ for $N=1$,
$\epsilon = 0.1253$ for $N=2$, etc... In particular, the rightmost maximum at $\epsilon = 0.2527$ combined with the initial state  $|1\ra$
provides stable, perfect population transfer, as shown in Fig. \ref{example3},
with less intensities than the ones required in the first protocol for a good fidelity,
since the value of $\epsilon$ is relatively large now.
Compare the values $\overline{\Omega} = 2\pi \times 0.48$ MHz and $\overline{E}/\hbar = 2\pi \times 5.89$ MHz for $\epsilon=0.2527$ in Protocol $3$ (with fidelity $F=1$)
by using Eqs. (\ref{energy-0}) and (\ref{energy-1}), with
$\overline{\Omega} = 2\pi \times 0.62$ MHz and $\overline{E}/\hbar = 2\pi \times 9.56$ MHz for $\epsilon=0.2$ (corresponding to $F=0.9682$) in the first protocol.
To achieve higher fidelity in the first protocol,
for example, $F=0.9998$, the time-averaged frequency and energy cost have to be increased up to $\overline{\Omega} = 2\pi \times 6.25$ MHz
and $\overline{E}/\hbar = 2\pi \times 981.49$ MHz by choosing $\epsilon=0.02$.
In summary, Protocol $3$ based on multi-mode driving provides an alternative shortcut to implement a stable, perfect population transfer
with a low energy cost.

%

%
%
%
\section{Discussions and Conclusions}
\label{summary}
In this paper, we have developed the invariant-based inverse engineering method to achieve fast population transfers in a three-level system. Two different
single-mode protocols are applied first in which the fidelity is linked to the laser intensity. Shortening the time also implies an energy cost.
Interestingly, to achieve the same fidelity, less intensity is required when the intermediate level $|2\ra$ is populated. A variant of the first protocol in which the initial state is simply the bare state $|1\ra$ and the dynamics is driven by a multi-mode wave-function provides a stable and less costly shortcut. Further exploration of the
multi-mode approach in this and other systems is left for a separate study.
The population of the intermediate level is usually problematic when its time decay
scale is smaller than the process time. While this may be a serious drawback for an adiabatic
slow process, it need not be for a fast shortcut. Protocols that populate level 2 may thus be considered as useful alternatives for sufficiently short process times.


As we stated in the Introduction, different techniques to find shortcuts to adiabaticity are strongly related, or even equivalent. The invariant-based inverse method presented here may be compared to the optimal control approach used in \cite{Boscain}. In
the optimal control method \cite{Boscain}, 
the system of control differential equations are the same as Eqs. (\ref{gamma}) and (\ref{beta}) in the invariant method.
The ultimate reason is that these equations are in fact equivalent to the Schr\"{o}dinger equation for a given wave-function parameterization.
The invariant dynamics provide thus a complementary understanding
of the optimal control approach, whereas optimal-control techniques
also help to optimize the results given by the invariant-based inverse
engineering.

Finally, the present results --within the on-resonance conditions-- are applicable to
quantum state transfer with three qubits \cite{Hu}, adiabatic splitting
or transport of atoms in a double well, and a triple well \cite{Corbalan}.
In a more general case, the Hamiltonian (\ref{gernal}) ($\Delta_p \neq 0$ and $\Delta_3 \neq 0$) does not possess SU(2) symmetry, so that
the invariant $I(t)$ should be constructed in terms of the eight Gell-Mann matrices for the SU(3) group \cite{Hioe}. The invariant-based inverse engineering for the systems with Gell-Mann dynamic symmetry will be discussed elsewhere.
\section*{Acknowledgement}
We acknowledge funding by the Basque government (Grant No. IT472-10),
Ministerio de Ciencia e Innovacion (Grant No.
FIS2009-12773-C02-01), and the UPV/EHU under program
UFI 11/55. X. C. also thanks
the National Natural Science Foundation of China (Grant No. 61176118).

\end{document}